A.K. Novokhrestov, A.D. Kalyakin, A.S. Kovalenko, V.S. Repkin

# Creating a vulnerable node based on the vulnerability «MS17-010»

The creation of a vulnerable node has been demonstrated through the analysis and implementation of the «MS17-010 (CVE-2017-0144)» vulnerability, affecting the SMBv1 protocol on various Windows operating systems. The principle and methodology of exploiting the vulnerability are described, with a formalized representation of the exploitation in the form of a Meta Attack Language (MAL) graph. Additionally, the attacker's implementation is outlined as the execution of an automated script in Python using the Metasploit Framework. Basic security measures for systems utilizing the SMBv1 protocol are provided.

In the modern context, one of the most critical issues in the field of information security (IS), along with the use of domestic software and hardware complexes, is the presence of skilled IS specialists. To fulfill their responsibilities, they must be familiar with current vulnerabilities affecting systems, as well as the threats that malicious actors may exploit.

For the effective training of specialists in new techniques and skills in the field of IS, the cyberpolygon Ampire has been created. Within its framework, learners are required to mitigate vulnerabilities exploited by automated code, simulating a real cybercriminal attack on an organization. The code runs on a previously created and prepared template, within which vulnerable nodes exist. A vulnerable node is one of the machines containing a software vulnerability that is exploited automatically using a script [1-4].

One of the main tasks for a cyber range is the continuous updating of the database of vulnerable nodes or complete scenarios in the form of interconnected nodes. That is why, within the scope of this article, the MS17-010 vulnerability was examined. A formal description of its exploitation variant was provided, and a vulnerable node based on it was created. The target for the attack was chosen as Windows 2012 Server, as there is still a high chance of encountering its usage in the corporate environment, combined with the outdated SMBv1 protocol.

**Vulnerability analysis of MS17-010**

The notoriety of the examined vulnerability «MS17-010» was brought about by a global hacker attack in 2017 named «EternalBlue», which was based on several vulnerabilities from the CVE list [5]. As can be understood, this vulnerability has been known for quite some time; it was added to the cve.mitre.org website on September 9, 2016, under the alternate name «Windows SMB RCE (Remote Code Execution) Vulnerability». It belongs to the family of vulnerabilities based on remote execution of arbitrary code on the victim's machine. Despite the «age» of the vulnerability, it can still be considered relevant, as RCE attacks are still actively utilized due to their high level of danger and effectiveness, providing malicious actors with complete control over the victim's machine.

The vulnerability «MS17-010» is associated with a flaw in processing an incorrectly crafted SMB (Server Message Block) request. The structure of SMB messages is documented and available on the Microsoft website [6].

Within MS17-010, nine different paths of vulnerability exploitation are considered. The most interesting among them are: «Wrong type assignment in SrvOs2FeaListSizeToNt()», leading to a buffer overflow, and «Transaction secondary can be used with any transaction type», which causes the server to not check the command sequence during SMB transaction execution, allowing the possibility of sending very large messages. Thus, for exploitation, it is necessary to have the ability to send transactional commands and have access to any «share» (IPC$ is quite suitable).

The ideal scenario for exploitation is a system with a version below Windows 8, as in this case, «anonymous (NULL) session» is available. This means that for successful exploitation, no additional knowledge about user accounts or «named pipes» [7] is required.

Vulnerable to this vulnerability is the SMBv1 protocol, which is used on the following operating systems:

– Windows Vista SP2;

– Windows Server 2008 (версий SP2 и R2 SP1);

– Windows 7 SP1;

– Windows 8.1;

– Windows Server 2012 (версий Gold и R2);

– Windows RT 8.1;

– Windows 10 (версий Gold, 1511 и 1607);

– Windows Server 2016.

For further implementation of the vulnerable node, the considered vulnerability was formally described in the form of a MAL (Meta Attack Language) graph.

**Software implementation of the attacker's actions**

Since the principle of the vulnerability has already been discussed above, further knowledge about exploitation was used to develop the vulnerable node.

To achieve this, two virtual machines were created in the VirtualBox virtualization environment, and the connection between them was properly configured. In this case, the first machine (Kali Linux), from which the attacker conducts the attack, has internet access and can reach servers located in the DMZ (Demilitarized Zone) of the company. The second machine (Windows Server 2012) is situated in the DMZ zone and has access both to the internet and to the internal network of the company, where typically data centers, development environments, management departments, etc., are deployed.

It is worth noting that in this case, the vulnerable node is only the machine running Windows Server 2012, on which the SMBv1 protocol is installed, representing the discussed vulnerability. The other machine is deployed and used for a comprehensive demonstration of the vulnerability exploitation and its consequences. The first machine is in the same subnet as the second, as the attacker's exploitation of this node is not considered the initial step. It is assumed that the attacker gained access to the DMZ earlier through the exploitation of another vulnerability.

To write automated code, it is necessary to use Metasploit. It is known that Metasploit is written in Ruby and does not support scripts written in Python. Despite this, Metasploit has a bidirectional RPC interface through which tasks can be executed.

There are two libraries that allow interaction with the remote procedure call (RPC) of Metasploit – pymetasploit by allfro and python-msfrpc by SpiderLabs [8]. In this case, the first library was used.

After installing this library, an RPC listener was launched, which will locally listen using the standard port «55553».

To connect from the attacker's machine to the previously started RPC listener in the final script, the following code is used:

*subprocess.Popen([«msfrpcd», «-P», «123», «-n», «-f», «-a», «127.0.0.1»], stdout=DEVNULL, stderr=STDOUT) time.sleep(10) client = MsfRpcClient(«123», port=55553, ssl=True)*

The script also utilizes the Metasploit framework, specifically the pre-built module «exploit/windows/smb/ms17_010_eternalblue», which contains the implementation of the MS17-010 exploit.

The code below represents the module declaration and the specification of the options required for the attack. In this case, only one option needs to be set – RHOST, which corresponds to the IP address of the target system.

*eternalblue = client.modules.use('exploit', 'windows/smb/ms17_010_eternalblue') eternalblue['RHOSTS'] = RHOST pl_eternalblue = client.modules.use('payload', 'windows/x64/meterpreter/reverse_tcp')*

The code then executes a loop with five iterations, where it checks for the existence of a created session and outputs messages about its creation using the command `run_module_with_output(eternalblue, pl_eternalblue)` (the function belongs to the pymetasploit3 library). If the output contains the strings «Session» and «Created in the background», then the exploit was successfully executed to create a Meterpreter session. In this case, the message «ms17_010_eternalblue exploit was used to create a Meterpreter connection!» is displayed, and the loop is terminated using the `break` statement.

If after the loop the variable `exploitation` remains False, the message «Exploitation unsuccessful!» is displayed. All the described actions are performed using the code provided below.

*for i in range(5):*

*exploitation = False if 'Session' and 'created in the background.'*

*In run_module_with_output(eternalblue, l_eternalblue):*

*exploitation = True print(«ms17_010_eternalblue*

*the exploit was used to create a Meterpreter connection!»)*

*break*

*if not exploitation:*

*print («Exploitation unsuccessful!»)*

As a result of executing this automated script, the attacker gains the ability to remotely execute arbitrary code on the Windows Server 2012 machine with System privileges. The attacker's access to the machine is depicted in Figure 1.

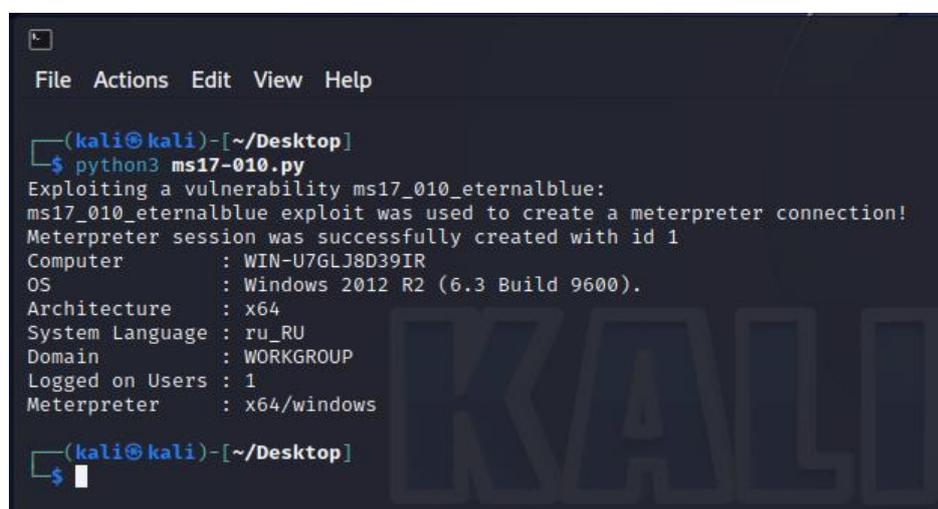

Figure 1. Information about the compromised machine

The consequences of this attack can include obtaining the complete configuration (and a list of protective mechanisms), leakage of information about users and their rights, password hash leakage (followed by brute-force attacks to recover passwords), leakage of Kerberos information (followed by password recovery and generation of silver tickets) – ultimately, a full compromise of the entire domain [8].

**Protection measures**

To limit the exploitation of the «MS17-010» vulnerability on systems with the SMBv1 server installed, the following steps can be taken:

- Update the Windows operating system to the latest versions, including the installation of security patches.

- Replace SMBv1, as it is an outdated and insecure version of the protocol for file and resource exchange in the Windows network. Use newer and more secure versions of the protocol, such as SMBv2 or SMBv3.

- Enable a firewall and network traffic filtering to restrict access to ports associated with SMB, securing the network devices that need protection.

- Utilize a network activity monitoring system within the internal network to identify suspicious behavior or unusual access attempts. This can be achieved using commercial antivirus programs, intrusion detection systems (IDS), and intrusion prevention systems (IPS).

**Conclusion**

As a result of analyzing the MS17-010 vulnerability, the mechanism and principle of its exploitation were examined. Additionally, the exploitation process within the vulnerable node was formally represented in the form of a Meta Attack Language graph (Figure 2). Formalizing the process in this case allowed for a specific and unambiguous definition of the sequential steps an attacker takes to exploit the vulnerability.

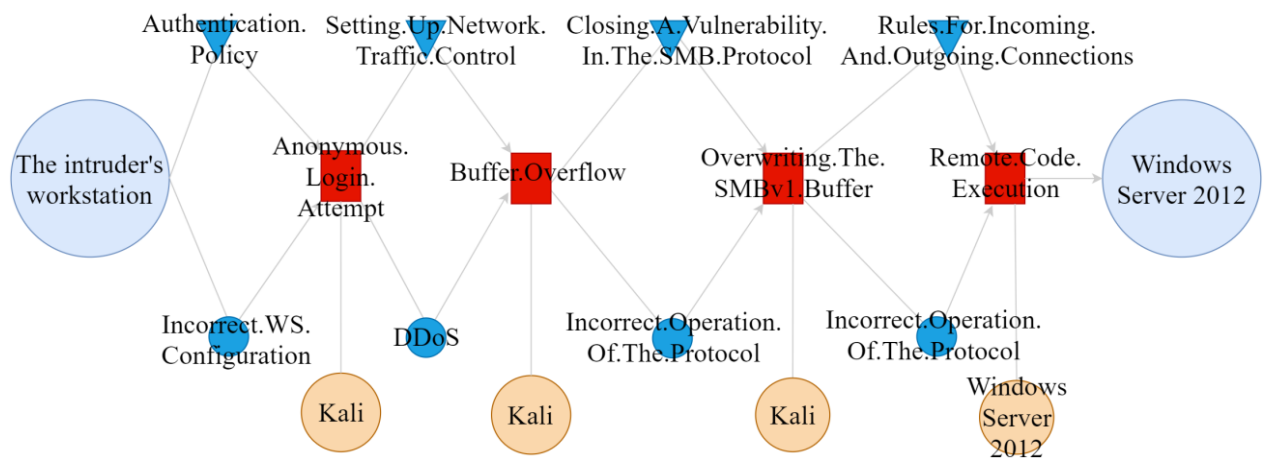

Figure 2. Graph using the Meta Attack Language

Using the acquired knowledge and the Meta Attack Language (MAL) graph, a vulnerable node was developed, incorporating a configured virtual machine with Windows Server 2012 and Python programming language code that simulates the actions of a real attacker exploiting the vulnerability using the Metasploit framework. In this form, the code and virtual machine seamlessly integrate into the cyberpolygon Ampire.

Additionally, some protective measures against the exploitation of this vulnerability were provided, which can aid in its elimination within the cyberpolygon Ampire or in a real-world operational scenario.

The work was carried out with the financial support of the Ministry of Science and Higher Education of the Russian Federation as part of the basic part of TUSUR University's state assignment for 2023–2025 (project No. FEWM-2023-0015).